\documentclass{iaus}
\usepackage{graphicx}

  \checkfont{eurm10}
  \iffontfound
    \IfFileExists{upmath.sty}
      {\typeout{^^JFound AMS Euler Roman fonts on the system,
                   using the 'upmath' package.^^J}%
       \usepackage{upmath}}
      {\typeout{^^JFound AMS Euler Roman fonts on the system, but you
                   dont seem to have the}%
       \typeout{'upmath' package installed. iaus.cls can take advantage
                 of these fonts,^^Jif you use 'upmath' package.^^J}%
      }
  \else
  \fi

  \checkfont{msam10}
  \iffontfound
    \IfFileExists{amssymb.sty}
      {\typeout{^^JFound AMS Symbol fonts on the system, using the
                'amssymb' package.^^J}%
       \usepackage{amssymb}%
         \let\leq=\leqslant
         
      }{}
  \fi

  \IfFileExists{amsbsy.sty}
    {\typeout{^^JFound the 'amsbsy' package on the system, using it.^^J}%
     \usepackage{amsbsy}}
    {}

\newsavebox{\astrutbox}
\sbox{\astrutbox}{\rule[-5pt]{0pt}{20pt}}

\title[Multiwavelength Survey by Yale-Chile]{MUSYC}

\author[P. Lira]%
{P. Lira$^1$ \and the MUSYC collaboration (www.astro.yale.edu/MUSYC/)}

\affiliation{
$^1$ Universidad de Chile, Casilla 36-D, Santiago, Chile\\
}

\pubyear{2004}
\volume{222}
\pagerange{1--8}
\date{?? and in revised form ??}
\jname{The Interplay among Black Holes, Stars and ISM \\in Galactic Nuclei}
\editors{Th. Storchi Bergmann, L.C. Ho \& H.R. Schmitt, eds.}
\begin{document}

\maketitle

\begin{abstract}

We describe MUSYC, a 1 square degree multiwavelength survey that will
make unique contributions in several areas and is particularly well
suited for the study of high redshift AGN.

\end{abstract}

\firstsection 

\section{Introduction}

Active Galactic Nuclei are signposts for supermassive black holes (BH)
and, because they are visible at very high redshifts, allow us to
probe the earliest collapsed objects.

Furthermore, galaxy formation -- specifically, the collapse of the
bulge and the initial burst of star formation -- appears to be
connected to the most intense phase of BH accretion (Kormendy \&
Gebhardt 2001). Essentially all galaxies with a bulge harbor
supermassive BHs (Gebhardt et~al. 2000, Kormendy \& Gebhardt 2001),
making the study of high-redshift AGN important for understanding
galaxy formation as well as the BH--galaxy connection. Different
scenarios have been proposed that could explain the observed link: the
simultaneous hierarchical growth of galaxies and their central black
holes through mergers (Haehnelt \& Kauffmann 2000) or a strong
coupling between BH accretion and star formation in proto-disks at high
redshift (Burkert \& Silk 2001).

Yet a potentially large fraction of AGN are not detected in the
traditional blue-excess or broad-emission-line surveys. Now in the new
era of deep surveys, it is possible to find all the AGN, whether
obscured or not. Combining multiwavelength data, especially in the IR
and X-rays, enables a complete census of accreting supermassive BHs,
as well as the measurement of the redshift distribution and evolution
of the full population.

\section{Survey Design}

A number of multiwavelength surveys aim to address these scientific
goals by exploiting the unprecedented combination of ground-based
observatories and space missions available today. Very deep,
pencil-beam IR/optical/X-ray surveys such as GOODS will probe the
faint end of the luminosity function of AGN. However, they are subject
to cosmic variance and the probability of actually observing objects
as rare as $z\sim6$ AGN is extremely low. Shallow, wide surveys such
as NOAO Deep/Wide, EIS Deep Public Survey, the Las Campanas Infra-Red
Survey (LCIRS), and ESO's K20 survey will cover larger areas to higher
flux limits in order to find large samples of AGN. Going too wide,
however, makes it difficult to obtain full coverage of high-quality
multiwavelength imaging, particularly in the near-infrared (JHK),
which is critically important for studying obscured AGN and where
detectors are smaller than optical CCDs.

The MUSYC (Multiwavelength Survey by Yale-Chile) project is a public
survey that comprises about 1 square degree of sky imaged to AB depths
of U,B,V,R=26.5 and K(AB)=22.5 for a $5 \sigma$ point source
detection, and is unique in its combination of depth and total
area. The survey covers four fields of $\sim
30^{\prime}\times30^{\prime}$ each, chosen to leverage existing data
and to enable flexible scheduling of observing time during the
year. Each field will be imaged from the ground in the optical and
near-infrared, and with space-based observatories in the optical (ACS
on HST), X-rays (Chandra, XMM), and mid/far-infrared (Spitzer).
Follow-up spectroscopy will be done mostly with multi-object
spectrographs (VIMOS, IMACS, GMOS). The survey fields will be a
natural choice for future observations with ALMA.

Fields were chosen to meet the following criteria: (a) Previous (long)
observations with space telescopes; (b) Low 100 micron emission,
reddening, and N$_{\rm H}$ columns; (c) No (or few) bright sources
known in the optical/radio; (d) High Galactic latitude ($|b| >
30^{\circ}$) to reduce stellar density; (e) Southern accessibility
($\delta \leq 5^\circ$). The chosen fields are the Extended CDF-South,
Extended HDF-South (see Fig.~1), SDSS 1030+05 (around a z=6.3 SDSS
quasar), and the 1256+01 `Castander Window'.

\begin{figure}
\centering
\includegraphics[scale=0.5]{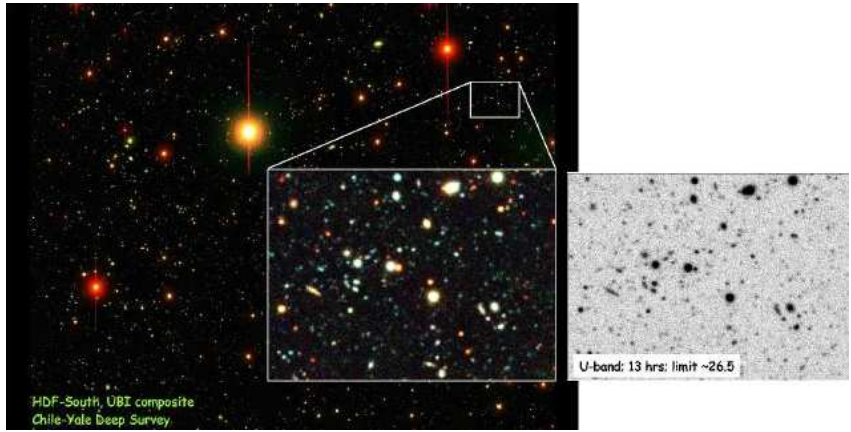}
\caption{UBI image of the $34^{\prime}\times34^{\prime}$ Extended
HDF-S. A zoom on a $3'\times 2'$ region shows a UBI composite and the
U-band image which is the deepest public wide-field U-band data
available."}
\end{figure}

\section{AGNs in MUSYC}

The main goal is to set constraints on the fraction of obscured AGN at
the $z\sim2$ QSO era. Understanding the evolution of BHs in the early
universe requires an increased sample of AGN at $z>1$. We will look
into the interplay between the growth of galaxies and their embedded
BHs by correlating the occurrence of AGN with host morphologies. We
will also estimate the cosmic growth of BH mass by converting AGN
luminosity to accreted mass. AGN are particularly likely to be
obscured by dust in a torus or warped disk around the central black
hole, but very few obscured AGN have been found. We will determine if
obscured AGN are actually rare or if they have been missed due to
biases in previous surveys. An unbiased AGN survey requires
multiwavelength data of the precise sort provided by MUSYC; the best
approach is to combine deep infrared and hard X-ray surveys with
optical or NIR spectra to measure redshifts.


\begin{thebibliography}{}

\bibitem[]{}
{Gebhardt, K. et~al.} 2000
\textit{ApJ} \textbf{539}, 13

\bibitem[]{}
{Kormendy J., Gebhardt K.} 2001
In \textit{20th Texas Symp. on Relativistic Astrophysics}
(ed. Wheeler, J. C. \& Martel, H.) pp. 363

\bibitem[]{}
{Burkert, A. \& Silk, J.} 2001
\textit{ApJ} \textbf{554}, L151

\bibitem[]{}
{Haehnelt, M. G. \& Kauffmann, G.} 2000
\textit{MNRAS} \textbf{318}, L35

\end{thebibliography}
\end{document}